\documentstyle[11pt,aspconf]{article}
\begin{document}

\vspace*{-2.3cm}
\begin{footnotesize}\baselineskip 10 pt\noindent
Proc. {\it Astronomical Data Analysis Software and Systems -- VI},~~Charlottesville,
Sept.~1996 \\
\hspace*{2cm} ASP Conference Series~~~~~\,(1997)~~~~~~~~~,~~eds.~G.\,Hunt et al.\\
\end{footnotesize}
\vspace*{0.7cm}

\title{
The CATS database to operate with astrophysical catalogs}
\author{O.V.~Verkhodanov and S.A.Trushkin}
\affil{Special Astrophysical Observatory, Nizhnij Arkhyz,
        Karachaj-Cherkessia, Russia, 357147}
\author{H.~Andernach}
\affil{INSA; ESA IUE Observatory, Apdo. 50727, E--28080 Madrid, Spain}
\author{V.N.~Chernenkov}
\affil{Special Astrophysical Observatory, Nizhnij Arkhyz,
        Russia, 357147}

\begin{abstract}
       A public database of astrophysical (radio and other) catalogs (CATS),
     has been created at Special Astrophysical Observatory (SAO).
It allows to execute a number of operations in batch or interactive mode,
e.g. to obtain a list and parameters of catalogs,
     to extract objects from one or several catalogs by various
     selection criteria,
  perform cross-identification of different catalogs,
      or construct radio spectra of selected sources.
     Access to CATS is provided in both dialog mode (non-graphics), and
     graphics mode (hypertext, via Tcl/Tk or possibly JAVA in future).
     The result of CATS operation can be sent to the user in
     tabular and graphical formats.
\end{abstract}


\section{Introduction}

%   Modern astrophysics operates with source parameters obtained
%in different spectral wavelength ranges with the goal to obtain a global
%understanding of physical properties and radiation processes
%in these objects. An ability of using different
%catalogs makes this problem considerably simpler. Lately
Different attempts have been made for combining many astronomical catalogs
in unified databases NED, SIMBAD, ESIS, ADS, etc.\,(see reviews by 
Andernach et al.\,1994; Andernach 1995).
Important shortcomings of these databases are~~(a) the incompleteness of the 
information stored compared to that offered in the original publication
and~~(b) the necessity to copy the whole catalog (if available at all) 
for dedicated work with it.
    We propose a new solution to this problem with a
``CATalogs supporting System'' (CATS) at
Special Astrophysical Observatory (SAO). It
allows to operate with catalogs coded in plain ASCII,
to cross-identify different radio catalogs,
to calculate spectral indices, to construct and fit spectra.
This database is now running on the server {\it ratan.sao.ru}\ of 
SAO of Russian Academy of Sciences (RAS),
and constitutes part of the data bank of SAO RAS
(Kononov, 1995).

\section {Realization of the database}

       The present active database of catalogs
of version 1.0 (Fig.1) is a unification of catalogs,
descriptions of catalogs and programs operating with them under
the freely distributed UNIX version OS Linux at the RATAN--600 server
with a Pentium processor (Verkhodanov \& Trushkin 1995).
The programs are coded in "C"  and are freely shared except for 
commercial goals.

\begin{figure}
%\centerline{\vbox{\psfig{figure=adass6_fig1.ps,width=12cm}}}
%\epsscale{.50}
%\plotone{catspic.ps}
\plotfiddle{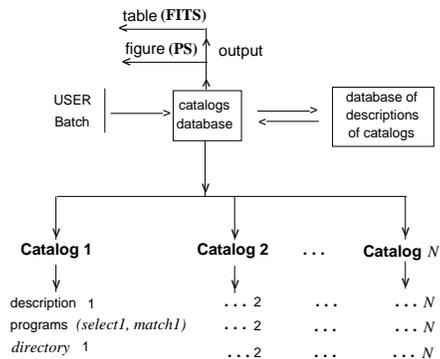}{5cm}{0}{45}{45}{-130}{-150}
\caption {
The scheme of the CATS database.
}
\end{figure}

New catalogs may be added to CATS in conformity with the following rules:
%\begin{enumerate}
%\item
1)
       every new catalog of objects
       has to be contained in the UNIX directory having the same name
       as the catalog of objects;
%\item
2)
       the description of the catalog
       must also be in that directory;
%\item
3)
       the programs for local operations with the catalog
       of objects are also in the same directory;
%\item
4)
       brief characteristics and names of the programs
       and file with the description of the catalog are stored
       in the file {\tt cats\_descr}.
%\end{enumerate}
%
% The catalog description file ({\tt cats\_descr}) contains
% the name and type of the catalog (radio, optical, mixed, etc.),
% frequency range, minimum fluxes or magnitudes, type of coordinates
% (equatorial or/and galactic), names of the local programs
% for the {\tt select} and {\tt match} functions, name of the
% document file, number of records in the catalog, resolution,
% calibration factor for obtaining the common scale of flux densities,
% and the reference.
%
% A user operates with CATS from any directory
% since the programs for general use of the catalogs are
% accessible system-wide.
%
The described manner of catalog storage eases the database
development, its expansion with new data and the fine-tuning
of the supporting programs.

Virtually all catalogs have a different format and list different
observables. It will be a major challenge to provide uniform access to such 
a heterogeneous collection of data sets based on different methods, using 
different notations and units (in the absence of a ``standard'' to create 
catalogs). Except for parameter-dependent derived quantities, we intend
to use all different fields (i.e. columns of data tables) as they were 
published.
%
% The concept of a ``Reference Directory'' (a central repository of
% metadata like the descriptions of fields for each catalog, their
% physical units, mapping of original field names to the actual name in
% the database, etc), would be extremely important to unify the ``look''
% of the catalogs as well as to process user queries.

% Only a clear understanding of the observing and data-processing techniques
% underlying a given catalog will guarantee reliable cross-IDs between
% catalogs. Therefore, documentation files in an agreed format will
% be prepared for all catalogs accessible to the search.
% Many catalogs provide only source names and will have to be completed
% with positional data. The latter may often be drawn from other databases
% (e.g.\ NED, SIMBAD, etc.), but eventually the DBMS will have to provide
% a ``name resolver'' to cope with the multitude of designations, which were
% especially diverse in the history of radio astronomy. For many other tables
% positional errors will have to be folded in (from formulae given in the
% publication)
% in order to minimize the amount of spurious cross--identifications.

\section {Capabilities and Access}
CATS is capable to accomplish the following tasks:
%\begin{enumerate}
%\item
1)
      Provide a short description and characteristics
      of each catalog; print the full list of catalogs
      relevant for a given sky area.
%\item
2)
      Select objects from one or several catalogs matching 
user-specified criteria, like equatorial and galactic coordinates, 
flux densities and spectral indices, observing frequency,
names of catalogs unified in mixed catalogs as Dixon's Master Source List, 
and object type (if provided by the catalog).
%\item
3)
      Cross--identification of different catalogs and selection
      after calculation of spectral indices.
%\item
4)
      Drawing radio spectra of selected sources in PostScript.
%\end{enumerate}

%\noindent
The result of the object selection can be sent to
the standard output or saved in the following formats:
%\begin{enumerate}
%\item
1)
      The original format of the input catalog.
%\item
2)
      A standard format, common for all catalogs and used further
      for unification and operation with spectra or other parameters.
      The standard FITS TABLE header describing the various
      data fields of the table may be recorded with
      the resulting file.
%\item
3)
      X-window codes or tape archives (TAR format) of compressed Post\-Script
      files for graphical spectra of radio sources.
%\end{enumerate}
%
%\noindent
The result of the CATS operation is an 'ASCII'-file
sorted according to different object characteristics. It can be used
for subsequent investigation of the radio source spectra or
statistical source properties in the RATAN--600
data processing system (Verkhodanov, 1997).

%\noindent
On-line access to CATS will be provided in several modes:
%\begin{enumerate}
%\item
1)
       Dialog mode (non-graphics) is the only mode established
       until now. Several UNIX {\it shell} scripts 
       (Verkhodanov \& Trushkin 1995) permit to run
       the database-supporting programs via TCP/IP and NFS
       protocols in the local computer net of SAO.
%\item
2)
       Access via TCL/Tk scripts on the basis of {\it shell}.
       This mode
       will also allow to operate with the figures or
       profiles of radio spectra or statistical distributions
       by different parameters of selected samples.
%\item
3)
       Hypertext access will eventually allow remote Internet users
       to operate with CATS via hypertext transfer protocol (HTML).
       It will allow to execute all described operations in graphic mode
       and probably take advantage of the JAVA language.
%\end{enumerate}

%\noindent
Presently we are working on a further type of access to CATS
by e-mail request enabling the user to send a message with his/her
requests.  The latter will be read automatically
and sent for execution to the CATS scripts. The result will be
sent automatically to the user via e-mail. 
Very bulky results will be placed in a public FTP area and
the user will be informed about the FTP address of the file(s).
The e-mail messages may have several formats describing
the search window, the coordinates of the center, the epoch,
the type of output format, and the type of catalogs in which to search.
CATS also permits to copy entire catalogs (via FTP) to the user's local
computer and to operate with them ``at home''.

\begin{figure}
%\centerline{\vbox{\psfig{figure=adass6_fig2.ps,width=12cm,angle=-90}}}
%\epsscale{.80}
%\plotone{spg.ps}
%\plotfiddle{spg.ps}{5cm}{-90}{30}{30}{-160}{180}
\plotfiddle{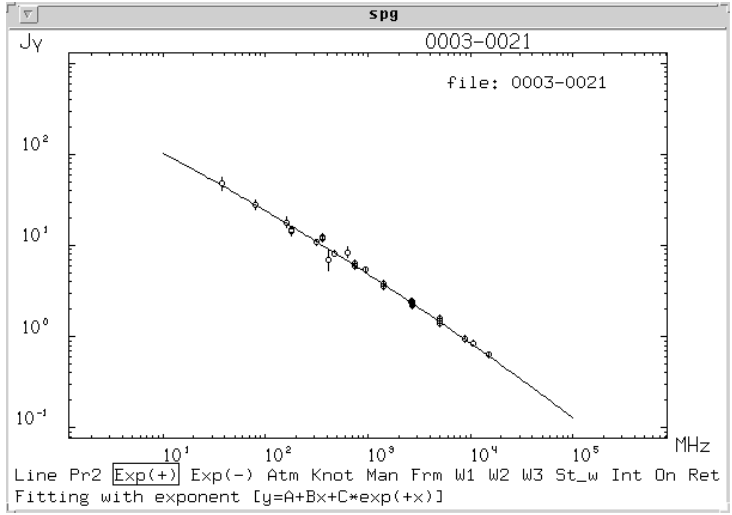}{6.8cm}{-90}{45}{45}{-195}{230}
\caption {
The {\it spg} screen with the menu and the radio spectrum.
}
\end{figure}
CATS operates with continuum spectra of radio sources recorded as FITS 
tables. The graphics program {\it spg} (SPectra Graphics) allows to fit 
a spectrum with a standard set of curves:
%\begin{itemize}
1) $y = A + Bx$,
2) $y = A + Bx + Cx^2$,
3) $y = A + Bx + C \times exp (x)$,
4) $y = A + Bx + C \times exp (-x)$,
%\end{itemize}
where $x = log~\nu$, $y = log~S$, $\nu$ is the frequency (MHz), $S$ is
the flux density (Jy).

The menu of the {\it spg} program (Fig. 2) allows to choose among these curves 
either automatically (by a least-squares fit), or by manual
selection of the fitting function, or by manual fitting using
the mouse (when the curve follows a cursor).
Data points may be weighted in different manners: setting
equal weight, setting weights by flux density errors or
filling a form with a table of frequencies, flux densities and weights.

\section {Conclusions}

    The development of CATS will provide a simple and 
convenient access to astrophysical information and
accelerate the process of obtaining characteristics of
celestial objects. Operation with the database will permit
astronomers to search for peculiar objects and study physical
processes in sources of cosmic radiation.

CATS allows not only to get accurate positions of radio sources
and to study radio spectra, but also to derive different
statistical properties of object samples.
Trushkin \& Verkhodanov (1995) recently demonstrated such possibilies 
in a cross--identification of two large catalogs in two different frequency 
ranges: the IRAS Point Source Catalog
in the infrared  and the UTRAO survey at 365 MHz.
Two of us (H.A. and O.V.) are working on the cross--identification and
eventual optical identification of UTR survey sources detected at 12--25 MHz.
We plan to solve the problem of very large error boxes by a stepwise
cross-ID progressing from low frequencies and angular resolution to
higher ones until the error box size permits the optical ID.

CATS is being expanded continuously. Presently it comprises over
50 catalogs including all the RATAN--600 catalogs
and occupies $\sim$250~Mb.  The database system
could be an essential part of a bigger project of the first publicly 
accessible database of radio sources, proposed by Andernach et al.\,(1997).

\acknowledgments
    This work is supported by the Russian Foundation of Basic
Research (grant No 96-07-89075).
    O.Verkhodanov thanks ISF-LOGOVAZ Foundation for the travel grant,
the SOC for the financial aid in the living expenses and
the LOC for the hospitality.

\end{document}